\documentclass{llncs}
\usepackage{amssymb, amsmath, amsfonts, stmaryrd}
\usepackage{url}
\usepackage{caption}
\usepackage{subcaption}
\pdfoutput=1

\usepackage{times}
\usepackage{microtype}
\usepackage{flushend}

\usepackage[svgnames,table]{xcolor}

\usepackage[textwidth=0.8in, 
            textsize=small, 
            color=Gold,            
            bordercolor=Gold]{todonotes}
\usepackage{listings}
\usepackage{xspace}
\usepackage{verbatim} 

\usepackage{hyperref} 
\hypersetup{
	colorlinks=true,
	urlcolor=black,
	linkcolor=black,
	citecolor=black,
	pdfstartview=FitH 
	}

\usepackage{courier}

\newcommand{\eg}{\emph{e.g.}\xspace}

\usepackage{cleveref}

\Crefname{section}{Sec.}{Sec.}
\Crefname{figure}{Fig.}{Fig.}
\Crefname{table}{Tab.}{Tab.}

\newcommand{\toolname}{Varanus}

\usepackage{pifont}

\newcommand{\ck}{\ding{51}}

\newcommand{\ofpp}{OpenFlow\ensuremath{^+}\xspace} 
\newcommand{\event}[1]{\ensuremath{\pi_{#1}}}
\newcommand{\timeof}[1]{\ensuremath{\tau_{#1}}}
\newcommand{\suffix}[1]{\ensuremath{\text{suffix}_{#1}}}
\newcommand{\env}{\ensuremath{\eta}\xspace}
\newcommand{\qryterms}{\ensuremath{R}\xspace}
\newcommand{\ltm}{\ensuremath{l_\tau}\xspace}
\newcommand{\lev}{\ensuremath{l_\pi}\xspace}
\newcommand{\envset}[1]{\ensuremath{H(#1)}\xspace}
\newcommand{\ticktock}{\ensuremath{\mathit{TICK}}\xspace}

\newcommand{\queryaut}{\ensuremath{\mathcal{A}_Q}\xspace}
\newcommand{\Nat}{\ensuremath{\mathbb{N}}\xspace}
\newcommand{\envhat}{\ensuremath{\widehat{\env}}\xspace}

\lstdefinelanguage{query}{
sensitive=false,
morekeywords={},              
alsoletter={\,,.,=},
keywordstyle=\bfseries\ttfamily,
captionpos=b,
frame=lrtb,
keepspaces=true,
float=false,
identifierstyle=\ttfamily,
mathescape=true,
basicstyle=\scriptsize\ttfamily,
showstringspaces=false,
firstnumber=1,
numberstyle=\tiny,
stepnumber=1,
breaklines=true,
numbers=left,
numbersep=5pt,
xleftmargin=8pt,
escapechar=*
}
\newcommand{\qry}[1]{\lstinline[language=query]{#1}}

\begin{document}

\title{Compiling Stateful Network Properties\\for Runtime Verification}

\author{
Tim Nelson\inst{1} Nicholas DeMarinis\inst{1} Timothy Adam Hoff\inst{1}\\ 
Rodrigo Fonseca\inst{1} Shriram Krishnamurthi\inst{1}
}
\institute{Brown University%\\
%\email{\{tn, ndemarin, thoff, rfonseca, sk\}@cs.brown.edu}
}

\pagestyle{plain}

\maketitle

\begin{abstract}
  Networks are difficult to configure correctly, and tricky to
  debug. These problems are accentuated by temporal and stateful
  behavior. Static verification, while useful, is ineffectual for
  detecting behavioral deviations induced by hardware faults, security
  failures, and so on, so dynamic property monitoring is also
  valuable. Unfortunately, existing monitoring and
  runtime verification for networks largely focuses on properties
  about individual packets (such as connectivity) or requires a
  digest of all network events be sent to a server,
  incurring enormous cost.

  We present a network monitoring system that avoids these
  problems. Because traces of network events correspond well to
  temporal logic, we use a subset of Metric First-Order Temporal Logic
  as the query language. These queries are compiled down to execute
  completely on the network switches. This vastly reduces network
  load, improves the precision of queries, and decreases detection
  latency. We show the practical feasibility of our work by extending
  a widely-used software switch and deploying it on networks. Our work
  also suggests improvements to network instruction sets to better
  support temporal monitoring.

\end{abstract}

\section{Introduction}
\label{sec:intro}

Network configurations are challenging to get right, and can be difficult~\cite{OGP03,wool:config-error-trends-cheese} to debug. 
Not only do they dictate static forwarding and packet-filtering behavior, 
they also control dynamic, stateful features such as
shortest-path routing, network-address translation, caching for various
protocols, and more. Given the ubiquity and criticality of networks, 
operators must be able to develop confidence in their configuration.

The difficulty is compounded in Software Defined Networks (SDNs), where operators
can write their own controller programs to govern switch
behavior. Not only is network behavior now defined by \emph{programs} rather
than \emph{configurations} (which tend to be of limited expressive power and
thus more amenable to reasoning~\cite{BBOR:end-to-end08,fogel++:nsdi15-batfish,kazemian++:nsdi13-netplumber,lg:firewall-queries09,mai:sigcomm11-anteater,n++:lisa-margrave-firewalls}), but the
mechanics of SDN itself induce new types of bugs~\cite{kuzniar++:flowtables-pam15,peresini++:ofcpp-hotsdn13} 
that must be addressed.

Static network-property checking tools like VeriFlow~\cite{khurshid:nsdi13-veriflow} 
and NetPlumber~\cite{kazemian++:nsdi13-netplumber}
are powerful, but limited in scope to to stateless forwarding policies.
Static tools designed specifically for SDNs will also function weakly at best
in a hybrid network that uses both SDN and traditional elements. Factors like
mis-connected cables~\cite{wordpress-config-error10}, external attacks,
and complex hardware problems~\cite{kuzniar++:flowtables-pam15} further demonstrate that 
static tools are complementary to and must be supported by strong runtime checks.

Purpose-built middleboxes, such as intrusion-detection systems,
monitor and react to traffic efficiently---often at line rate or
close to it. But these are costly, lack general-purpose flexibility,
and sit at least one step removed from the forwarding performed by
the switches.  High-performance Network Function Virtualization
(NFV) solutions (\eg~\cite{martins++:clickos-nsdi14,palkar++:e2-sosp15}) are now able to process
traffic at or near line rate (10Gpbs), but still require traffic to
be rerouted through them. Another class of monitoring tools, including
NetSight~\cite{handigol++:netsight-nsdi14} and
Simon~\cite{nylfk:sosr15-simon}, send digests of packets seen to a
central server---an approach that adds flexibility while incurring
significant cost.

Narayana, et al.~\cite{narayana++:path-queries-nsdi16} introduce a different technique: compiling
regular-expression \emph{path queries} down to rules executed by switches. However, path queries 
are limited to expressing how a single packet traverses the network, independent of history.
History is vital for monitoring the stateful behavior of protocols and applications.
A central server might support stateful queries by updating stateless rules on each switch as needed,
but this approach increases latency, server load, and risk that the monitor will be impacted by
reordering, retransmission, or other eccentricities of the network.

Instead, we present \toolname, a novel system for \emph{stateful runtime verification of network-control programs}:
compiling temporal queries \emph{in their entirety} to the switches themselves.
Our switch-centric capture approach has several key advantages. It:
\begin{itemize}
	\item reduces load on the controller and/or monitoring server; 
	
	\item decreases latency of detection and state-transition, which 
	also lessens the potential for unsound results (\Cref{subsec:complete});
	
	\item avoids potential event re-ordering induced by transmission to a central server; and
	
	\item increases the expressive power of queries by exposing how the switch
	transforms an arriving packet into a departing packet, without resorting
	to unsound methods such as comparing payload-hashes~\cite{handigol++:netsight-nsdi14}.
	
\end{itemize}

This approach is challenging because switches are optimized to match and forward packets
as quickly as possible, limiting the expressive power of rules they can execute. 
In fact, current standard switch-rule formats
are entirely stateless, making temporal-query compilation unworkable. Nevertheless, switches do have non-trivial
computational resources: a slower, more expressive tier of processing sits above its fast, limited matching capabilities
in order to handle rule-updates from the controller and other operations that cannot be performed at line rate.
\toolname{} takes advantage of this, conservatively extending switch rules to support a constrained subset 
of Metric Temporal Logic, retaining fast match capabilities at the lower tier while performing state updates 
in the higher tier---balancing expressiveness with performance.
We demonstrate feasibility by modifying a popular software switch and deploying queries in a network.

\section{Queries: Examples, Syntax, Semantics, and Expressive Power}
\label{sec:language}

Before detailing \toolname{}'s query syntax and semantics, we first motivate temporal 
queries with a small example: monitoring a stateful-firewall SDN application. 

\subsection{Queries by Example}
\label{sec:examples}

A stateful firewall blocks packets that arrive on external ports unless 
they are part of a connection initiated from the internal network. 
Compared to most network applications, a stateful firewall is relatively simple, yet
its implementation could contain many bugs: outgoing traffic or valid
incoming traffic could be rejected (resulting in a connectivity failure), or
invalid external traffic could be permitted (a security threat). 
For simplicity, we limit the firewall switch to two ports: internal and external.
We also introduce a black-listed external host with network address \qry{192.0.2.1}. We now have a number of correctness properties; for instance: 
\begin{enumerate}
	\item Packets from host \qry{192.0.2.1} never egress the firewall. 
	\item Packets arriving at an internal port are always forwarded to the external port.
	\item After a packet from internal host A to external host B arrives at an internal port, packets from B to A arriving
	at an external port will be output on the internal port for the next ten seconds.
	\item If no packet from internal host A to external host B has previous arrived at an internal port, packets from B to A
	at an external port will be dropped.
\end{enumerate}
We wish to monitor the network for counterexamples, i.e., traces of 
network events that falsify each property.

\paragraph{Property 1}
If the switch never modifies the IP source field, then this property can be monitored 
by a single stateless packet filter that checks for blacklisted packets departing from 
the firewall. The corresponding query is:
\begin{lstlisting}[label=lst:property0,language=query]
see p: egress | p.nwSrc = 192.0.2.1       
\end{lstlisting}
which notifies the monitor when the switch observes a packet egress with source network address \qry{192.0.2.1}---either to raise an alert or to trigger further processing.

\paragraph{Property 2}
In contrast to the prior property, this one requires state: the system only incurs an obligation
to forward a packet once it has arrived. Detecting when this property fails means looking for 
cases when a packet egresses in an unexpected way (e.g., it is dropped).
Thus, we now have a sequence of two observations:
\begin{lstlisting}[label=lst:property1,language=query]
see p:  arrival           | p.locPt   = 1             // *{\rmfamily Internal port}*
see p': egress same       | p'.locPt != 2             // *{\rmfamily External port}*
\end{lstlisting}
These lines should be read with an implicit ``followed by'' between them.
The identifier \qry{p} matches an arbitrary packet arrival at the internal port ({\tt 1}). 
Next, \qry{p'} matches when the packet is not seen departing the external port ({\tt 2}).
The \qry{same} annotation specifies that \qry{p'} must be an egress for 
the packet identified in the previous arrival observation (\qry{p}).
The second observation depends on the first; a given egress only 
discharges the obligation incurred by its preceding arrival.

\paragraph{Property 3}
The obligation in this property is only incurred after \emph{multiple} distinct packets:
\begin{lstlisting}[label=lst:property2,language=query]
see p1: arrival            | p1.locPt = 1
see p2: arrival within 10  | p2.locPt = 2, p2.nwSrc = p1.nwDst, 
                             p2.nwDst = p1.nwSrc
see p2': egress same       | p2'.locPt != 1 
\end{lstlisting}
The identifier \qry{p1} matches a packet arriving at the internal port, which should open a hole in the firewall for return traffic. 
\qry{p2} is a corresponding return packet arriving on the external port \qry{within} 10 seconds of \qry{p1}.
The observation for \qry{p2'} matches when \qry{p2} does not depart on port 1,
meaning it was not forwarded to the internal network.

\paragraph{Property 4}
Here, a counterexample must contain \emph{no} analogue to \qry{p1} 
in the above:
\begin{lstlisting}[label=lst:property3,language=query]
not see p1              | p1.locPt = 1, p2.nwSrc = p1.nwDst, p2.nwDst = p1.nwSrc
  until see p2: arrival | p2.locPt = 2
see p2': egress same    | p2'.locPt = 1
\end{lstlisting}
The first observation (lines 1--2) matches an incoming packet \qry{p2} on the external port, 
provided that no packet \qry{p1} (bearing reversed addresses) has previously arrived
at the internal port. \qry{p2'} then observes that \qry{p2} 
has erroneously been forwarded.

Even these short examples require non-trivial features: recognizing chains of
related observations, time-dependence, detecting when two events involve the same packet, 
negative observations, awareness of prior events, etc.

\subsection{Syntax}
\label{subsec:syntax}

A query comprises an ordered list of \emph{observations} 
$\lbrack O_1, ..., O_n\rbrack$, each of which matches a network event 
(a \emph{positive} observation) or lack of an event within some period of time 
(a \emph{negative} observation). Each observation has a \emph{type}: 
either a packet arrival or packet egress (possibly with temporal constraints).

Each observation also contains a \emph{trigger event} identifier and a \emph{match predicate} which 
references that identifier and (optionally) identifiers of prior positive observations.
Also, \qry{until} observations contain a \emph{blocking event} identifier and a corresponding \emph{blocking predicate}. Each predicate consists of a set of literals.
\Cref{fig:syntax} gives the complete grammar.

\begin{figure}[t]
\begin{lstlisting}[label=lst:syntax,language=query,numbers=none]
QUERY ::= LISTOF(OBSERVATION)
OBSERVATION ::=
  "see" ID ":" TYPE "|" PREDICATE |
  "not see" ID ":" TYPE "|" PREDICATE |
  "not see" ID "|" PREDICATE  
    "until" see ID ":" TYPE "|" PREDICATE |
PREDICATE ::= LISTOF(ATOMIC)
ATOMIC ::= TERM "=" TERM | TERM "!=" TERM
TERM ::= NUM | ID.FIELD 
TYPE ::= "arrival" | "arrival within" NUM | 
         "egress" | "egress within" NUM | "egress same"
FIELD ::= dlTyp | dlSrc | dlDst | nwSrc | nwDst | nwProto | ...
\end{lstlisting}
\caption{\small Query syntax.
Non-terminals are capitalized, and {\tt LISTOF(X)} denotes a list of expressions matching
non-terminal {\tt X}. {\tt FIELD} matches any valid packet field name.}
\label{fig:syntax}
\hrule
\end{figure}

To be \emph{well-formed}, a query $\lbrack O_1, ..., O_n\rbrack$ must also satisfy the following:
\begin{itemize}
\item every negative observation has a \qry{within} annotation;
\item the \qry{same} type annotation only occurs with \qry{egress};
\item every \qry{egress same} observation is preceded by a positive \qry{arrival} observation;
\item each $O_i$, $O_j$ ($i \neq j$) are labeled by different identifiers; and
\item each $O_i$'s predicate refers only to its own identifier and (optionally) 
identifiers of prior observations $O_j$ ($j \leq i$). 
\end{itemize}
These ensure key semantic requirements, e.g., every term is bound before it is used.

\subsection{Semantics}
\label{subsec:semantics}

Queries match traces of events in the network. Observations within a query
match individual events or timeouts in the context of a trace.
To make this precise, we first introduce some notation.

An \emph{event} is one of: a packet arrival, a packet egress, a clock tick (written as \ticktock), or the no-op event ($\bot$) 
which the semantics uses as a placeholder when no event has yet occurred. 
A \emph{timestamp} is a non-negative integer value.
A \emph{trace} is a finite sequence of $($event, timestamp$)$ pairs with monotonically non-decreasing timestamps. 
For a trace $T$, we use $\event{i}(T)$ to refer to the $i^{th}$
event in a trace and $\timeof{i}(T)$ to refer to its timestamp. $\suffix{k}(T)$ denotes the suffix of $T$ obtained by removing the leading $k$ elements. We omit the application ``$(T)$''
where the trace is clear from context. 

Each query has a fixed set of \emph{query terms} \qryterms: the \qry{e.fld} expressions that occur within its observations. 
Since there are finitely many observations in a query, and only finitely many valid field names, the set of query terms is always finite.
To project out individual field values from events, we use a partial function $getf: Events \times FIELD \rightarrow NUM$. 
If an event $e$ does not contain a field \qry{f}, then $getf(e, f)$ is undefined.
We use $O(e)$ to indicate an observation whose match identifier is $e$; for an \qry{until} observation, 
this is the identifier in the second position, i.e., the trigger.
We assume traces contain a discrete tick event for each positive integer $i$,
denoted $(\ticktock, i)$; this is for convenience in the formalism for 
observations using \qry{within}, which in practice (\Cref{sec:impl}) will be implemented via switch-rule timeouts.

For every query $Q$ and trace $T$, we now define whether $T$ \emph{satisfies} $Q$
(written as $T \models Q$). Since $Q$ is a sequence of observations $\lbrack O_1, ..., O_n \rbrack$, 
we must also define whether $T$ satisfies each observation $O_i$. But whether $T \models O_i \,(1 \leq i \leq n)$ may depend on
when and how \emph{previous} observations were satisfied: $O_i$ may reference field values
bound by $O_1, ..., O_{i-1}$, and may also, via \qry{within} keyword, depend on the time since the last observation was satisfied.
We provide this past-time context via an environment \env,
which maps syntactic query terms to concrete values such as packet fields. For instance, \env may
map \qry{p1.nwDst} (the network destination address of the event bound to \qry{p1}) to \qry{10.0.0.1}.

\Cref{fig:semantics} defines the meaning of a query via 
a logical satisfaction ($\models$) relation, similarly to temporal logic. 
Since this semantics introduces numerous parameters, for brevity we write 
$T \models Q$ for $T \models_{\emptyset, 0, \bot} Q$, omitting the initial environment,
last-match timestamp, and last-match event.

\begin{figure*}[th]
\centering
\begin{tabular}{|l l l|}
\hline
\multicolumn{3}{|c|}{Atomic Formulas ($\mathbf{m}$) and Predicates ({\bf C}) match an event with respect to an environment.} \\
\hline  
  $\event{} \models_\eta \text{\tt e.f = c}$       & $\iff$   & $getf(\event{}, f) = c$ \\  
  $\event{} \models_\eta \text{\tt e.f1 = e.f2}$   & $\iff$   & $getf(\event{}, f1) = getf(\event{}, f2)$ \\
  $\event{} \models_\eta \text{\tt e.f1 = t}$      & $\iff$   & $getf(\event{}, f1) = \eta(t)$ \\
  (if \texttt{t} a term that is neither \texttt{c} nor \texttt{e.f2}) & & \\
  $\event{} \models_\eta \text{\tt t1 != t2}$      & $\iff$   & $\event{} \not\models_\eta \text{\tt t1 = t2}$ \\
\hline
  $\event{} \models_\eta \mathtt{m_1, ..., m_k}$  & $\iff$   & $\forall m_i . \event{} \models_\eta m_i$ \\  
\hline
\end{tabular}

\medskip

\begin{tabular}{|l l l|}
\hline
\multicolumn{3}{|c|}{Event Types ($\mathbf{Y}$) match a trace position with respect to an environment,} \\
\multicolumn{3}{|c|}{a last-match timestamp, a last-match event, and an operator ($\leq$ or $=$). } \\
\hline
  $T, i \models_{\eta, \ltm, \lev, op} \text{\tt arrival}$          & $\iff$   & $\event{i}$ is a packet arrival \\  
  $T, i \models_{\eta, \ltm, \lev, op} \text{\tt egress}$           & $\iff$   & $\event{i}$ is a packet egress \\  
  $T, i \models_{\eta, \ltm, \lev, op} \text{\tt arrival within D}$ & $\iff$   & $\event{i}$ is a packet arrival, and $\timeof{i} \,op\, (\ltm+D)$ \\  
  $T, i \models_{\eta, \ltm, \lev, op} \text{\tt egress within D}$  & $\iff$   & $\event{i}$ is a packet egress, and $\timeof{i} \,op\, (\ltm+D)$ \\  
  $T, i \models_{\eta, \ltm, \lev, op} \text{\tt egress same}$      & $\iff$   & $\event{i}$ is a packet egress, and $I(\lev, \event{i})$ \\    
\hline
\end{tabular}

\medskip

\begin{tabular}{|l l l l|}
\hline
\multicolumn{4}{|c|}{Observations ($\mathbf{O}$) match a trace position with respect to an environment,} \\
\multicolumn{4}{|c|}{a last-match timestamp, and a last-match event.} \\

\hline
  $T, i \models_{\eta, \ltm, \lev}$ & $\text{\tt see e : Y | C}$            & $\iff$ & $T, i \models_{\eta, \ltm, \lev, \leq} Y$ and $\event{i} \models_\eta C$ \\

\hline
  $T, i \models_{\eta, \ltm, \lev}$ & $\text{\tt not see e : Y | C}$        & $\iff$ & $T, i \models_{\eta, \ltm, \lev,=} Y$, $\event{i} = \ticktock,  \text{ and }$ \\
                                    &                                       &        & $\forall j (0 \leq j < i), \event{j} \not\models_\eta C$ \\    

\hline
  $T, i \models_{\eta, \ltm, \lev}$ & $\text{\tt not see e1 | C1}$          &        & $T, i \models_{\eta, \ltm, \lev, \leq} Y$, $\event{i} \models_\eta C2 \text{ and }$  \\
                                   & $\;\text{\tt until see e2 : Y | C2}$   & $\iff$ & $\forall j (0 \leq j < i)$, either \\ 
                                   &                                        &        & $T, j \not\models_{\eta, \ltm, \lev, \leq} Y$ or $\event{j} \not\models_{\eta+[e2, \event{j}]} C1$ \\
\hline
\end{tabular}

\medskip

\begin{tabular}{|l l l|}
\hline
\multicolumn{3}{|c|}{Queries ($\mathbf{Q}$) match traces with respect to an environment, a last timestamp, and a last event.} \\
\hline

  $T \models_{\eta, \ltm, \lev} \mathtt{O_1(\mathtt{e_1}), ..., O_n}$ & $\iff$   & $\exists i \geq 0 \,|\, T,i \models_{\eta, \ltm, \lev} \mathtt{O_1}, \text{ and }$ \\
           & & $\suffix{i+1} \models_{\eta+[e_1, \event{i}],\timeof{i},\event{i}} \mathtt{O_2, ..., O_n}$\\    
\hline
\end{tabular}
\caption{\small Logical Semantics for {\bf Q}ueries, {\bf O}bservations, event t{\bf y}pes, predi{\bf c}ates and atomic for{\bf m}ulas.
	We use $\eta+[e, \event{}]$ as shorthand for $\eta[\forall f \in \mathtt{FIELD},\; \mathtt{e.fld} \mapsto getf(\event{i}, f)]$,
    i.e., environment $\eta$ extended by the fields of $\event{}$ under identifier \qry{e}. 
    The symmetric binary relation on events, $I$, represents whether two events involve the same packet.
}
\label{fig:semantics}
\hrule
\end{figure*}

\subsection{Expressive Power}
\label{subsec:logic}

Since \toolname{}'s language can match sequences of events, it has a clear expressive 
advantage over existing work that focuses only on stateless monitoring. 
Implementations of real world protocols, such as ARP and DHCP, as well as bespoke SDN controller programs
require a stateful monitoring language.
While more expressive monitoring solutions exist, such as Simon~\cite{nylfk:sosr15-simon},
these approaches are not conducive to switch-based state management. (\Cref{sec:relwork} discusses other work in more detail.) 
Our language provides a novel middle-ground: meeting the expressive
demands of stateful monitoring while also achieving the benefits of switch-based capture.

\paragraph{Comparison to Temporal Logic}
Our query language comprises a strict subset of Metric First-Order Temporal Logic (MFOTL)~\cite{basin++:mfoltl-jacm15} 
that is conducive to execution on switches.
In essence, positive observations correspond to $F$ (``finally'') or $U$ 
(``until'') formulas, and negative observations to $G$ (``globally'')
formulas, with later observations nesting. Forward references in \qry{until} 
observations can be expressed in MFOTL using existential variables. For instance, \framebox{\qry{not see e1 | e1.nwSrc = e2.nwDst until see e2: arrival}}
corresponds to $\exists x .\, \neg(src = x) \,U_{[0,\inf)}\, (dst = x)$.
The fixed temporal structure of queries---observation followed by observation---leads to restricted formula 
nesting in the equivalent MFOTL, easing compilation while sacrificing expressive power. For example, there is no
equivalent query for the MFOTL formula $F \alpha \wedge F \beta$.
Also, while the \qry{within} keyword allows observations to be satisfied at any point within 
the interval $[0, D]$, MFOTL allows both ends of intervals to be specified. This
limitation makes it possible to compile \qry{within} via switch rule timeouts (\Cref{sec:compile}).

\paragraph{Additional Examples}

To demonstrate the broad applicability of this technique, \Cref{fig:more-examples} lists additional stateful properties for five example applications not seen in \Cref{sec:examples}: 
(1) a learning switch, which learns how to reach new destinations as it observes traffic; 
(2) a Network Address Translation (NAT) application, which masquerades multiple hosts as a single address; 
(3) an Address Resolution Protocol cache, which facilitates fast lookup between layer 2 and layer 3 addresses; 
(4) a packet-tagging application; and
(5) port knocking, which allows traffic from external hosts that send a pre-specified packet sequence.
Each of these safety properties can be expressed as a query in \toolname{}.

\definecolor{gray1}{gray}{0.95}

\noindent
\begin{figure*}[t]
\rowcolors{1}{gray!25}{white}	
\begin{tabular}{|c|l|c|c|c|}
	\hline
	\rowcolor{gray!50}
	App                   & Property                                                       & N? & IE?  & U? \\
	\hline
	\hline
	\cellcolor{gray!50} Learning              & Never flood after destination learned                          &    &    & \\
	\cellcolor{gray!50} Switch                & Once destination learned, send out correct port                &    &    & \\
	\hline 
	\hline        	
    \cellcolor{gray!50} NAT                   & NAT ports used consistently once assigned                      &    &\ck &  \\ 
	\hline
	\hline
	\cellcolor{gray!50} ARP                   & No new requests generated for cached addresses                 &    &    & \\                                                      
	\cellcolor{gray!50} Cache                 & Requests for cached addresses always replied to                &\ck &    & \\
 	\cellcolor{gray!50}                       & If no prior request received, propagated on core ports         &\ck &    &\ck \\ 
	\hline
	\hline	
	\cellcolor{gray!50} VLAN                  & Once a tag is seen on a port, never see another                &    &    &\\
	\cellcolor{gray!50} and misc.             & See at most $k$ distinct tags on any port                      &    &    & \\
	\cellcolor{gray!50} tagging               & Untagged packets always tagged before departure on trunk       &    &\ck &\\                                    
    \cellcolor{gray!50}                   	  & Tagged packets always de-tagged before departure on access port&    &    &\\                                                          
	\hline
	\hline
	\cellcolor{gray!50} Port Knocking         & Recognize correct knocking sequence                            &    &    &\ck\\
    \cellcolor{gray!50}                       & Knocking packets should be dropped                             &    &\ck & \\
	\hline                          
\end{tabular}
\caption{\small Select additional stateful queries, beyond those in \Cref{sec:examples}, which fit our language. Columns 1 and 2 name the specific program and property. Additional columns contain a \ck\ if the query uses negative observations ({\bf N?}), ingress-egress pairing ({\bf IE?}), or until observations ({\bf U?}).}
\label{fig:more-examples}
\hrule
\end{figure*}

\section{Queries as Automata}
\label{sec:automata}

We now show how a query can be realized as a non-deterministic automaton over
event traces. This fact will be useful for compilation:
in \Cref{sec:compile} we will see that
each state of these automata can be represented as a set of switch rules.

Intuitively, a state of such an automaton carries several things: (1) which observation
was last matched; (2) the current environment; (3) a set of forbidden environments;
(4) a last-match event; and (5) a last-match timestamp. Much of this state-construction
echoes the semantics in \Cref{sec:language}; the only exception is (3), which allows the
automaton to remember previously seen events that matched the blocking part 
of an \qry{until} observation.

Given a query $Q = \lbrack O_1, ..., O_n \rbrack$, let \envset{\qryterms} be the
set of all environments over the query terms $\qryterms$ of $Q$ and let $|Q|$
be the set $\{ O_1, ..., Q_n\}$.
The \emph{query automaton} for $Q$, denoted by $\queryaut$,
is the non-deterministic automaton
$(S, \Sigma, \delta, s_0, F)$ where:

\begin{enumerate}
\item the set of states $S$ is $(|Q| \times \envset{\qryterms} \times 2^{\envset{\qryterms}} \times \mathit{Event} \times \Nat) \cup \{(O_0, \emptyset, \emptyset, \bot, 0)\}$ (where $O_0$ is a distinct value representing ``no observations satisfied'');
\item $\Sigma$ is a set of possible $($event, timestamp$)$ trace elements;
\item $\delta$ is a relation on $S \times \Sigma \times S$ (the \emph{transition relation}); 
\item $s_0 = (O_0, \emptyset, \emptyset, \bot, 0)$ is the unique starting state; and
\item the set of accepting states $F = \{(o, \env, \envhat, \event{}, \timeof{}) | o = O_n \}$; i.e., all states for which the final observation has been matched.
\end{enumerate}

\begin{figure*}[t!]
\begin{minipage}[b]{0.25\linewidth}
\includegraphics[width=1.5in]{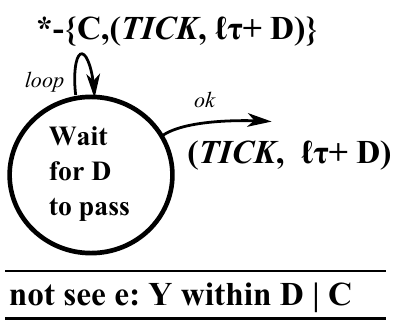}
\end{minipage} 
\hfill
\begin{minipage}[b]{0.68\linewidth}
\begin{align*}    
\delta_i          & = \delta_i^{loop} \cup \delta_i^{ok}   \\            
\delta_i^{loop}   & = \{ ((Q_i, \env, \envhat, \lev, \ltm), (\event{}, \timeof{}), (Q_i, \env, \envhat, \lev, \ltm)) \\
                  &  \text{ for every \env, \envhat, \lev, \ltm, \event{}, and \timeof{}} \\
                  & \text { unless $\event{} \models_\eta C$ or (\event{}=TICK \text{ and } $\timeof{} = \ltm+D$)} \\
\delta_i^{ok}     & = \{ ((Q_i, \env, \envhat, \lev, \ltm), (\event{}, \timeof{}), (Q_{i+1}, \env+[e, \event{}], \emptyset, \event{}, \timeof{})) \\
                  & \text{ where } \event{}=TICK \text{ and } \timeof{} = \ltm+D \}  \\
\end{align*}
\vspace{-12mm}
\end{minipage}

\caption{\small Producing the non-deterministic transition relation for a query $\lbrack O_1, ..., O_n \rbrack$.
The notation $A-[e]$ means the subset of match predicate $A$ with all literals involving
event $e$ removed. For space, we show only the \qry{not see within} case here; \Cref{app:automata} contains the full table.}
\label{fig:automata}
\hrule
\end{figure*}

If $\delta$ is undefined for a particular state and input, the 
automaton is said to \emph{halt} on that pair. The automaton \emph{accepts} 
a trace if there is some run on the trace that ends in an accepting state.
Given a finite-branching transition relation, every finite trace induces
only finitely many runs.
The transition relation $\delta$ is defined as $\delta = \bigcup_{1 \leq i < n} \delta_i$,
and each $\delta_i$ is defined for each observation $O_i$. \Cref{fig:automata} gives $\delta_i$ for 
one type of observation; \Cref{app:automata} provides the full definition. 
The automata produced here are quite limited: the only cycles they contain are self loops, 
and environments grow monotonically as each run evolves. Furthermore, $\delta$
is finite-branching. The only subtlety is in $\envhat$, which represents forbidden bindings accumulated
in the course of matching an \qry{until} observation. \Cref{fig:autrun} demonstrates a partial run of the automaton for Property 4 of \Cref{sec:examples}.
Query automata correspond closely to queries, as the following theorem (proven in \Cref{app:thm1}) shows:

\begin{figure*}[t!]
\includegraphics[width=4.75in]{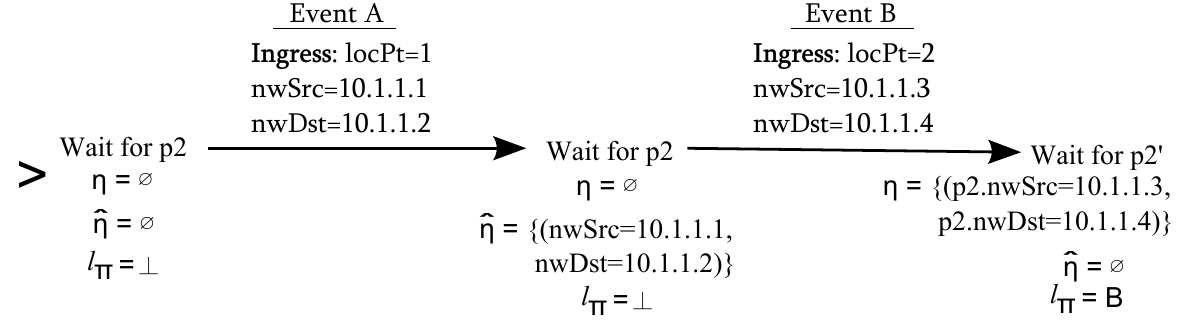}
\caption{\small Example run of the automaton for Property 4 (timer details omitted).
The starting state has empty $\env$ and $\envhat$ and waits for \qry{p2}. When a packet (A)
matching \qry{p1} arrives, the new state continues waiting for \qry{p2}, but has learned a
forbidden environment. Since (B) matches \qry{p2} but does not match the forbidden environment,
it advances the state to waiting for \qry{p2'}. If B egresses (not shown) at port 1, the
automaton advances to an accepting state.}
\label{fig:autrun}
\hrule
\end{figure*}

\begin{theorem}
Let $T$ be a trace and $Q = \lbrack O_1, ..., O_n \rbrack$ a query. Then $T \models Q \iff \queryaut$ accepts $T$.
\end{theorem}

\subsubsection*{The Role of Non-determinism in this Domain}
Non-determinism allows multiple runs of \queryaut\ to co-exist.
Consider the query:
\begin{lstlisting}[label=lst:need-nd1,language=query]
see p1: arrival | p1.nwSrc = 10.1.1.1
see p2: arrival | p2.nwDst = p1.nwDst\end{lstlisting}
executed over a trace containing packets with multiple destination addresses. Each destination
will produce a distinct environment binding \qry{p1.nwDst}, necessitating multiple
states be ``active'' at any one time. Timers also make heavy use of non-determinism, even 
for queries where no environment is needed. For instance:
\begin{lstlisting}[label=lst:need-nd2,language=query]
see p1: arrival          | p1.nwSrc = 10.1.1.1
see p2: arrival within 5 | p2.nwDst = 10.1.1.2 
see p3: arrival within 5 | p3.nwSrc = p3.nwDst 
\end{lstlisting}
This query is entirely positive, and requires no environment to 
evaluate---the constraints on \qry{p2} are not dependent on \qry{p1}. 
Suppose we execute it on the following event trace:
\begin{lstlisting}[label=lst:need-nd2-trace,language=query,numbers=none]
t=0:  nwSrc = 10.1.1.1,  nwDst = 10.1.1.5       [matches p1]
t=4:  nwSrc = 10.1.1.1,  nwDst = 10.1.1.2       [matches p1,p2]
t=8:  nwSrc = 10.1.1.10, nwDst = 10.1.1.2       [matches p2]
t=10: nwSrc = 10.1.1.3,  nwDst = 10.1.1.3       [matches p3]
\end{lstlisting}
Any individual run will time out after 5 seconds of waiting for \qry{p2}, but 
non-deterministic automata can have multiple simultaneous states. This allows the
wait for \qry{p2} to be extended via a new state at {\tt t=4}, enabling the capture of \qry{p2} at {\tt t=8}. 
(The logical semantics enable
this via ``for some $i$'' in the {\bf Q} portion of \Cref{fig:semantics}.) 
Rather than attempting to determinize query automata,
we will embrace non-determinism and use it explicitly in our compiler. 

\section{Compilation}
\label{sec:compile}

\toolname{} generates switch rules in an extension of OpenFlow~\cite{McKeown:ccr08-openflow},
a widely-used protocol that allows the controller to install persistent forwarding rules on switches.
The OpenFlow protocol is widely supported both by major hardware vendors, and by software switches such as
Open vSwitch~\cite{pfaff++:ovs-nsdi15}. 
Because of this extensive support, our compiler targets OpenFlow, 
although the approach is not OpenFlow specific.

The OpenFlow standard continues to evolve, and 
individual vendors also extend the protocol to provide new features.
Our work fits into this culture of extension, using monitoring to explore the 
benefits of carefully increasing the OpenFlow instruction set.

\subsection{Background}
\label{subsec:of-background}

In OpenFlow\footnote{``OpenFlow'' without a version modifier will henceforth refer to a simplified version of
OpenFlow 1.3, atop which we implement our changes.}, the \emph{rules} that dictate switch behavior are held
in ordered collections called \emph{tables}.
An OpenFlow rule is a tuple $(D, T, M, A)$ consisting of 
an optional hard timeout {\bf d}uration, 
a {\bf t}able identifier,
a set of {\bf m}atch criteria, 
and a set of {\bf a}ctions to perform if the rule is matched. 
Match criteria are predicates that match packet fields against values, possibly with wildcarding. For instance, a rule might
match packets with source IP address {\tt 10.1.1.*} and destination TCP port {\tt 80}. Actions include forwarding
the packet out a constant port ({\tt out(k)}), overwriting a packet field with a constant value ({\tt fld:=val}),
 resubmitting the packet to a constant table ({\tt resubmit(t)}), and 
sending the packet to the controller for further instructions ({\tt controller}). 
Rules also contain a numeric priority, which we abstract out and give implicitly via list ordering.

When a packet arrives, it is submitted to table $0$. Within each table, the highest-priority matching rule applies, 
and its actions are added to the overall \emph{action set}. The switch performs all actions in the set, in a fixed order 
(i.e., field modifications before output), once no further table resubmits are mandated. The graph of resubmits must be acyclic.
An empty action set results in a dropped packet. If the packet matches \emph{no} rule in any table, it is
sent to the controller. Crucially, this packet-evaluation process 
leverages the limited expressive power of rules to process packets quickly, whether
via specialized data-structures in software or optimized hardware. Only cache-misses and 
modifications to the rule-set need pass through a switch's slower general-purpose
datapath.

In effect, these simple static rules are a cache of controller instructions, which may be proactively installed. 
In standard OpenFlow,
only the controller, and not the switch itself, can modify or install rules. This provides a clear separation
of concerns between switch and controller, but leads to increased latency and load. Different dynamic-rule
extensions have been proposed, which we compare in \Cref{sec:relwork}. Open vSwitch also implements a dynamic extension: the ``learn'' action, which 
allows a rule, when matched, to install a new rule that does not itself use the learn action. Our own extensions build atop this.

\subsection{Rule Extensions with Dynamic Actions}
\label{subsec:ofpp}

\toolname{} extends the learn action to allow learned rules to themselves have learn actions, 
i.e., we enable nesting rules to be learned up to arbitrary bounded depth.
Concretely, we extend OpenFlow as follows:
an \emph{\ofpp} rule is a tuple $(D, T, M, A, O)$ where $O$ is an action set to perform on rule time{\bf o}ut.
Other components are identical to OpenFlow, except that the set of permitted actions in $A$ is larger, including
deleting all rules in the current table ({\tt delete\_all}), incrementing an atomic counter ({\tt inc(c)}), 
and \emph{learning} a new rule ({\tt learn(r)}). Table identifiers are split between
\emph{ingress} and \emph{egress} tables; egress tables apply immediately before the packet departs (or is 
dropped by) the switch.

\subsection{Compiling to Switches}
\label{subsec:compile}

To realize $\queryaut$ as a set of \ofpp rules,
\toolname{} produces an initial rule set that embodies the automaton's starting state.
It contains, via nested learn actions, a blueprint for all possible runs of $\queryaut$.
Since the switch accumulates an action set from \emph{all} tables before
discharging the packet, this
gives us a way to implement non-determinism: every table corresponds to 
an active run of $\queryaut$. Separate table spaces for ingress and egress
maintain the distinction between \qry{arrival} and \qry{egress} events.

Consider a run of $\queryaut$ in state $S = (O_i, \env, \envhat, \ltm, \lev)$. 
If $S$ is an accepting state, no rules are required; the monitor will have been
notified on transitioning to $S$.
If $S$ is not accepting, there are four varieties of transitions to cover:
{\bf (1)} true self-loops (defined by each $\delta^{loop}$ in \Cref{fig:automata}); 
{\bf (2)} halting; 
{\bf (3)} forward transitions from an $O_i$ state to a $O_{i+1}$ state ($\delta^{ok}$ in \Cref{fig:automata}); and 
{\bf (4)} in the case of \qry{until}, blocking self-loops that rule out potential forward
transitions but remain in a state with the same $O_i$ component ($\delta^{block}$ in \Cref{app:automata}'s \Cref{fig:automata-full}).
The compiler addresses these as follows:
\begin{enumerate}
\item[(1)] Self-loops are the default; if an event matches no rules, no action is taken. 
\item[(2)] A run halts in two cases: when a timed positive observation times out, or
a negative observation is matched. 
The compiler realizes these by adding a timeout to corresponding rules.
For positive observations, the timeout erases the rules;
for negative observations, forward-transitions execute on timeout and matching packets delete the rules.
Since tables correspond to runs of $\queryaut$, deleting a table removes that run. 
\item[(3)] Forward transitions use either the \qry{learn} action (to learn rules into a fresh table) or 
the \qry{controller} action (to notify the monitor that the query has been satisfied).
\item[(4)] To realize the blocking transitions for \qry{until} observations, the
compiler adds an extra, low-priority rule that, as blocking events are seen,
learns new rules to block corresponding packets from triggering forward transitions.
\end{enumerate}

The \toolname{} compiler recursively descends through each observation,
building a corresponding set of switch rules.
Where match predicates involve negative literals, the compiler inserts
\emph{shadow} rules before the \emph{trigger} rule 
that advances the state. Since only one rule per table
can apply, these block the trigger from applying.

Observations in a query can reference previously seen events, but switch rules cannot carry an environment directly.
To bridge this gap, we exploit techniques used for compiling the lambda calculus: substitution and De Bruijn indices~\cite{debruijn:distance72}.
The compiler embeds current values via substitution as new rules are learned. 
To prevent premature substitution, every field reference in switch rules also carries a \emph{deferral} value.
Where De Bruijn indices count how many bindings away a variable was bound, deferral values count the number of learn actions to wait before substituting. Our extensions to OVS 
read these deferral values when learning, perform substitution if it is zero, and otherwise decrement it in the new learned rule.
\Cref{app:alg} gives the algorithm in full.

\section{Implementation and Evaluation}
\label{sec:impl}

Our implementation comprises the compiler of \Cref{sec:compile} and a version of 
the Open vSwitch~\cite{pfaff++:ovs-nsdi15} software that has been modified to support \ofpp. We also
use a proxy between the switch and any controller applications
(such as the stateful firewall in \Cref{sec:examples}), allowing us receive query notifications and place any
application-installed rules between our compiled ingress and egress tables.
We optimize pairs of \qry{ingress}/\qry{egress same} observations by using switch registers to 
maintain ingress field values as needed; this allows us to avoid queuing packets until 
all learn actions are complete.

The full version of OpenFlow involves advanced features not mentioned above; 
we have discarded these for simplicity in our prototype. 
Consequently, our ability to detect egresses is limited
to packets emitted on zero, one, or all ports (i.e., not multicast).

\subsection*{Performance Evaluation}

The key performance metric in our approach is the time required to
match packets in our query tables, which directly affects forwarding
latency.  This additional latency includes the
time to execute our modified learn actions, and the time to
evaluate a packet in each active table.
Since the time to execute our modified learn actions does not
differ significantly from the time to execute a standard learn action,
we focus on the number of active tables.  
Critically, the number of tables affects \textit{all} packets
processed by Open
vSwitch's tables, not just those that match a query observation.  
To quantify the latency introduced by adding tables, we created a test
network using an installation of Open vSwitch with our extensions and
two hosts running in Mininet~\cite{lantz++:hotnets10-mininet}.

\begin{figure*}[t!]
\includegraphics[width=\columnwidth]{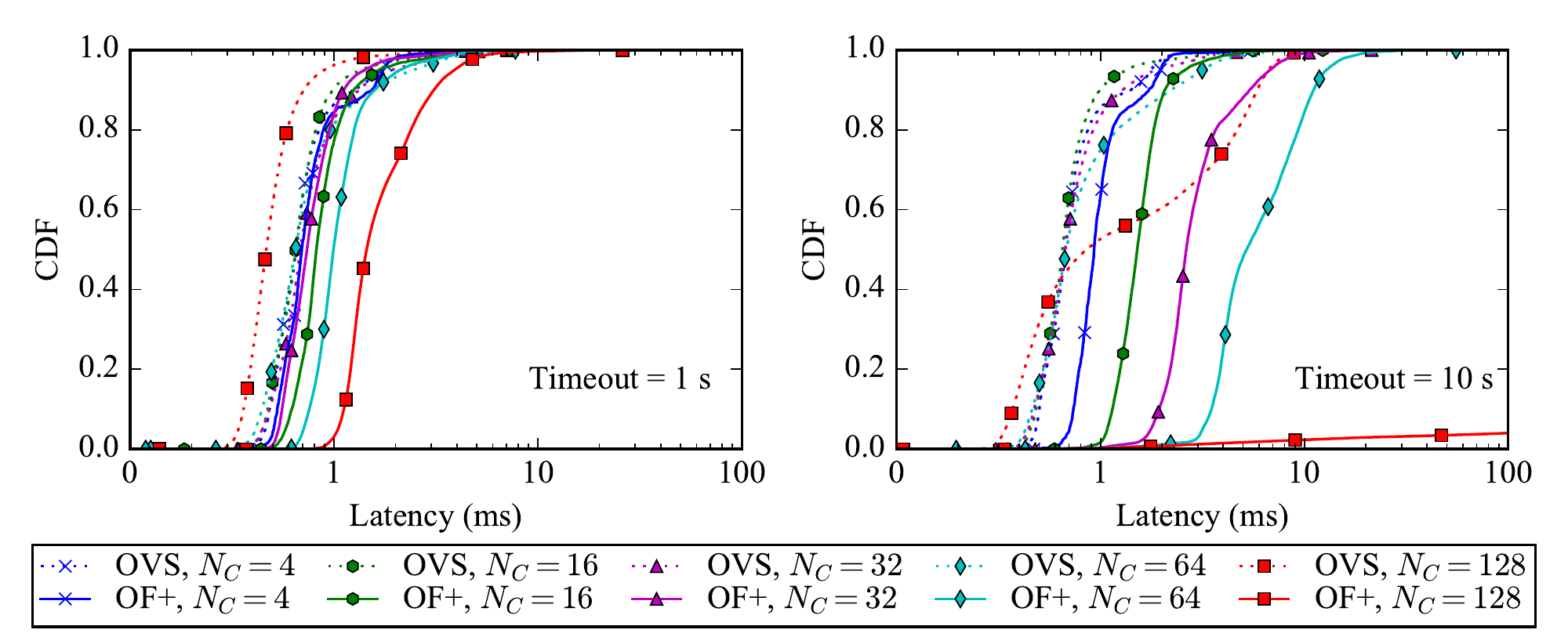}
\caption{\small CDF of latency for a base installation of Open vSwitch (OVS) and our
  modified switch with support for \ofpp rules (OF+).  Our evaluation
  measured the latency to process a new connection for a number of
 new connections per second ($N_C$) and with rule timeouts of 1 second
 (left) and 10 seconds (right), shown on a log scale.
Evaluating \ofpp rules at 128 connections/sec with 10
second timeouts (not fully pictured) shows prohibitively high latencies in the 1-10sec range.  
  }
\label{fig:latency}
\hrule
\end{figure*}

OVS maintains a low-level cache of actions 
for packets already processed by rule tables. In order to focus on the packets most
affected by our modifications, we manufacture a worst-case situation by creating a 
large number of fresh TCP connections, which result in cache misses, each of which 
causes all tables to be evaluated.
We measured the latency of rules that allocate a new table for each new TCP
connection entering the switch, similar to the stateful-firewall queries of \Cref{sec:examples}.
We generated traffic using {\tt ovs-benchmark}~\cite{pfaff++:ovs-nsdi15}, a
performance measurement tool included with Open vSwitch, and captured traffic with the {\tt tshark} packet-analysis tool.
We examined the round
trip time for the first two packets in the TCP handshake, which
is the portion of the connection that must be processed by the slow
path.  To understand how the latency changes based on the number of
active runs of the automaton, we varied the numbers of new connections
per second and the duration of the timeout on installed rules.  
(Without timeouts, the amount of query state retained increases monotonically as packets are processed, which
becomes intractable.)
For each set of parameters, we conducted measurements for 5
minutes of constant traffic at the specified rate, which we express as
a CDF in \Cref{fig:latency}.

To provide a baseline for the performance of the learn action, we
use an unmodified OVS installation learning
new rules into a \emph{single} table for each new TCP connection seen. In unmodified OVS, the learn action
has a latency of about 1 millisecond, due to the need to perform an expensive rule table
update for each packet.  This latency increases based on the
number of new connections and rules in the table, as shown in \Cref{fig:latency}.

The difference between the baseline implementation and \toolname{}
is reflected in \Cref{fig:latency} as the horizontal distance between 
OVS (dashed lines) and \ofpp\ (solid lines), and is due to the added
cost of maintaining and evaluating multiple \ofpp\ tables.
In our test network, traffic at 64 new connections per second with 1
second rule timeouts can be handled by about 500 tables (one for each of the 8 packets in each {\tt ovs-benchmark} TCP connection within the one-second window) and incurs a
latency under 1 ms. 128 new connections per second, requiring
1000 tables, are each handled within 10ms.  
For the 10 second timeout case, the latencies are much higher due to
the increased number of active tables; at 128
connections per second with 10 second timeouts, we reach the
resource limit of our test switch at approximately 8000 tables, which
presents as an unacceptably high latency (on the order of tens of
seconds), since uncached table evaluation cannot keep up with the new traffic.

These measurements indicate that \toolname{} can be practical on a small network;
we allow the user to select suitable timeout values for their queries, as they are in the best
position to know both their needs and what their environment can support.
\Cref{sec:conclusion} discusses future work on alternative approaches 
that improve scaling on larger networks.

\section{Related Work}
\label{sec:relwork}

Metric temporal logics have been used previously for runtime verification, 
notably by Basin, et al.~\cite{basin++:mfoltl-jacm15}, whose MonPoly tool monitors 
a richer subset of MFOTL formulas---including unrestricted nesting and two-sided intervals---than our approach.
Kim, et al.~\cite{kim++:mac-ecrs99} also give a runtime program monitoring framework, MaC, 
with a query language inspired by past-time LTL with intervals. Their language adds state via 
``auxiliary variables'', which are updated as a monitoring script progresses through an event stream. 
MaC's support for time intervals and aggregation is superior to ours, but as
auxiliary variables are limited to basic Java types (e.g., numbers), the
language cannot support branching caused by non-deterministic 
interleaving (\Cref{sec:automata}).
The Eagle~\cite{barringer:eagleltl-ipdps04} monitoring logic is also more expressive than ours, 
allowing the definition of new temporal operators. 
Deshmukh, et al.~\cite{deshmukh++:rv15-stl} perform off-line system monitoring using Signal Temporal Logic, which enables checking quantitative, rather than strictly Boolean, properties.
All these works assume a standard Turing-complete execution target. In contrast, \toolname{} integrates with existing
switch software to minimize the impact of on-line capture and stateful filtering. General monitoring tools also neglect
network-domain distinctions such as the connection between packet ingress and corresponding egress.

Simon~\cite{nylfk:sosr15-simon} provides a scriptable, 
interactive monitoring framework for SDN programs. 
However, Simon runs on a
centralized server, which means \emph{all packet events} on the
network must be communicated to this server, whereas our approach
reduces load and improves event granularity. Events generated by \toolname{} could be provided to Simon streams for further 
processing and interactive exploration.

Narayana, et al.~\cite{narayana++:path-queries-nsdi16}
compile ``path queries'' about packet forwarding to switches. 
Path queries, by their nature, are distributed across multiple switches, whereas we compile stateful queries only for single switches.
Path queries are restricted to regular expressions over 
possible forwarding paths, and so cannot express the array of stateful properties that \toolname{} targets, but do not require extensions to OpenFlow.

DMaC~\cite{zhou++:dmac-rv09} extends MaC to networking by
compiling MaC properties to Network Datalog~\cite{loo:commacm09-declarative-networking}.
DMaC can describe both stateful and path-specific properties, and can monitor behavior across multiple switches,
whereas our properties are single-switch.
DMaC requires that switch state and traffic information be made available in a form it can query,
and executes via a custom Datalog engine on software switches.
Since our goal is to filter for temporal packet sequences on optimized switch
software---with minimal impact on forwarding speed---our switch-rule based approach differs significantly.

NetSight~\cite{handigol++:netsight-nsdi14} captures packet trajectories
by capturing digests for each packet, which are processed off-switch. 
As our work focuses on compiling temporal, multi-packet queries to switches, it
is largely orthogonal, but our extensions to switch rules potentially provide NetSight 
with a means to reduce the number of packet digests produced.
Moreover, NetSight differentiates packets by hashing, whereas our approach allows a more
fine-grained distinction.

Production SDN monitoring fabrics, such as BigTap~\cite{bigtap}, are limited to operations 
supported by current versions of OpenFlow, and thus cannot avail themselves of switch-side stateful monitoring. 
Planck~\cite{rasley++:planck-sigcomm14} and other sampling approaches deliberately compromise on completeness in the face of intrinsic network constraints.
We do not sample deliberately, although overfull switch queues can manifest similarly (\Cref{subsec:complete}).

Numerous other tools (e.g., \cite{porras:hotsdn12-fortnox,agarwal++:sdn-traceroute-hotsdn14,zhu++:everflow-sigcomm15})
perform runtime monitoring, but either lack compilation to switches
or monitor only single-packet paths through the network.
Others (e.g.,~\cite{scott++:sts-causal-sigcomm14,viswanathan:data-analysis-nsdi11,wundsam++:ofrewind-atc11})
focus on data analysis after the fact, rather than real-time monitoring.

FAST~\cite{moshref:fast-hotsdn14},  OpenState~\cite{bianchi:openstate-ccr14}, and POF~\cite{song:pof-hotsdn13} conservatively extend the power of OpenFlow rules. 
In all of these, switches maintain additional ``per-flow'' state. A flow is defined by matching a set of header
fields, giving a fixed equivalence relation on traffic: equivalent packets use the same index into the state table.
While some of our queries could be compiled to these extensions,
in general our equivalence relation needs to vary with the current state---one observation may match on IP source address, 
another on TCP port.
FAST and OpenState both assume a deterministic automaton for state evolution, and so using them would prevent us from
leveraging non-determinism as we do. 
The P4~\cite{bosshart++:p4-ccr14} language also allows limited state on switches, in the form of persistent registers.
Our state is more self contained since rules can only learn fresh rules or delete their own table (\Cref{sec:compile}),
rather than affecting other rules---as is possible via shared registers.

Purely static tools, such as network configuration analyzers~\cite{BBOR:end-to-end08,lg:firewall-queries09,kazemian++:nsdi13-netplumber,mai:sigcomm11-anteater,fogel++:nsdi15-batfish,n++:lisa-margrave-firewalls} and languages built for analysis (e.g., \cite{kim++:kinetic-nsdi15,nfsk:nsdi14-flowlog})
are powerful but not sufficient for robust network testing.
Veriflow~\cite{khurshid:nsdi13-veriflow} statically analyzes
\emph{updates} to (stateless) OpenFlow switch rules in real time. This hybrid approach 
is nevertheless limited to switches governed by an SDN controller.  Since
our approach only \emph{collects events} on SDN switches, it can be used even in a hybrid network.
Beckett et al.~\cite{beckett++:assertion-debug-hotsdn14} enhance SDN programs with annotations that 
reference program state directly. These compile to new VeriFlow assertions as the program state changes, and are therefore limited to analyzing switch-rule updates.

\section{Discussion and Conclusion}
\label{sec:conclusion}

To our knowledge, \toolname{} is the first work to enable runtime verification of SDN programs by compiling 
cross-packet, temporal queries directly to switches. At the moment, we have focused mainly on query expressiveness, yet
have kept our extensions to OpenFlow narrow in order to limit detrimental effects on forwarding speed.

While some monitoring goals can be met by installing middleboxes at the network edge, runtime verification of
SDN applications---which govern the entire network---requires visibility into the network
core as well as its edges.  Moreover, testing and debugging are eased by the ability to quickly rewrite and deploy new queries to any 
\ofpp switch in the network---rather than reprovisioning capture hardware for each new requirement. Our approach
provides both of these features, lending it power, and flexibility compared to current methods.

Regardless of how queries are eventually deployed---on purely software switches running OVS
extensions or on hardware---it remains vital for switches to process
packets as quickly as possible, which is not feasible without limitations on
the expressive power of packet-matching. We respect this constraint by
producing rules with packet-matching criteria that echo existing fast-match 
instruction sets while also containing state-modification actions that can be
executed asynchronously.

\subsubsection{Soundness and Completeness}
\label{subsec:complete}
In an environment such as a switch, which lacks atomicity guarantees, 
the race between processing events and advancing state means that some amount of both unsoundness (false positives) and incompleteness (false negatives)
are unavoidable for stateful queries. Switches are optimized to forward traffic
rapidly (on the order of micro- or nanoseconds); rule update is relatively
glacial (whole milliseconds). Moreover, switches have limited queue space, and may drop packets
under heavy traffic conditions. 
To avoid these problems, one might imagine forcing the switch to process only one packet at a time.
While intuitively appealing, this is unacceptable in a network that handles even a modest amount of traffic.
These challenges are intrinsic to the domain, and while monitoring
directly on switches (as opposed to on a server) greatly
reduces network-induced inaccuracy, we leave the larger problem for future work.
Probabilistic approaches such as traffic sampling~\cite{rasley++:planck-sigcomm14} 
present an especially promising avenue.

\subsubsection{Scaling, Hardware, and Future Work}

As our evaluation shows, performance degrades as the number of active tables increases.
Our current approach installs each new run of $\queryaut$ as a new block of \ofpp rules 
in a fresh table. The system does this because, in general, a single
packet may advance multiple partially-completed runs. However, it appears that
full support for non-determinism is not always necessary. For instance, when falsifying
Property 3 (\Cref{sec:examples}), a single packet can never advance two 
partially-completed runs unless they are waiting for different observations.
We therefore believe that one table for each observation should suffice to capture the query,
and that this can be echoed in the formalism by partial determinization of $\queryaut$.
Using only a constant number of tables---where possible---would drastically improve
performance. Moreover, using a fixed number of tables makes pipelining possible, 
easing application of our techniques in purpose-built forwarding hardware.

\bibliographystyle{splncs03}
\bibliography{tn}

\appendix 
\newpage
\section{Automaton Construction}
\label{app:automata}

\begin{figure*}[ht!]
\begin{minipage}[b]{0.25\linewidth}
\includegraphics[width=1.5in]{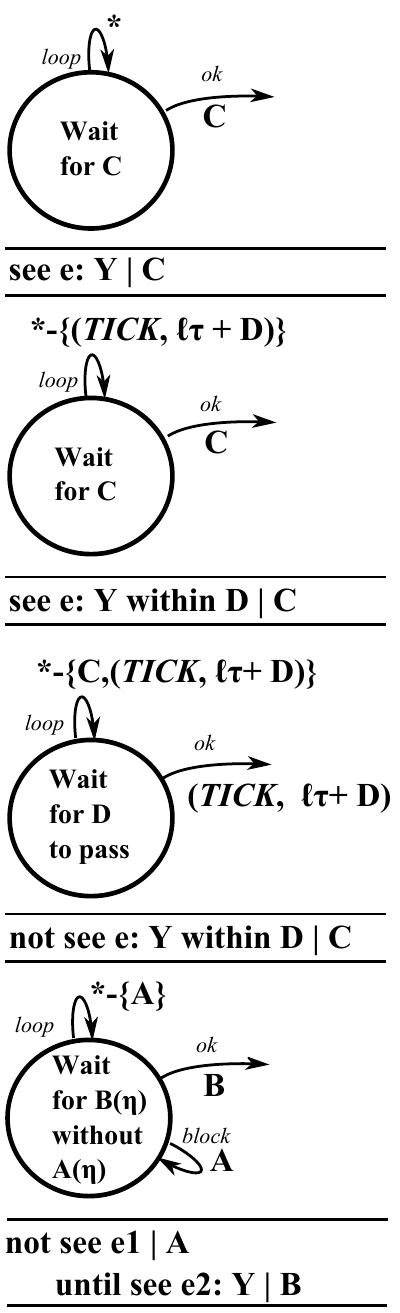}
\end{minipage} 
\hfill
\begin{minipage}[b]{0.68\linewidth}
\begin{align*} 
\delta_i^{loop}   & = \{ ((O_i, \env, \envhat, \lev, \ltm), (\event{}, \timeof{}), (O_i, \env, \envhat, \lev, \ltm)) \\
                  &  \text{ for every \env, \envhat, \lev, \ltm, \event{}, and \timeof{}} \} \\
\delta_i^{ok}     & = \{ ((Q_i, \env, \envhat, \lev, \ltm), (\event{}, \timeof{}), (Q_{i+1}, \env+[e,\event{}], \emptyset, \event{}, \timeof{})) \\
                  & \text{ where $\event{} \models_\eta C$} \} \\ 
\end{align*}
\vspace{-15mm}

\begin{align*} 
\delta_i^{loop}   &  = \{ ((Q_i, \env, \envhat, \lev, \ltm), (\event{}, \timeof{}), (Q_i, \env, \envhat, \lev, \ltm)) \\
                  &  \text{ for every \env, \envhat, \lev, \ltm, \event{}, and \timeof{}} \\
                  &  \text{ unless } \event{}=TICK \text{ and } \timeof{} = \ltm+D \}  \\
\delta_i^{ok}     & = \{ ((Q_i, \env, \envhat, \lev, \ltm), (\event{}, \timeof{}), (Q_{i+1}, \env+[e, \event{}], \emptyset, \event{}, \timeof{})) \\
                  & \text{ where $\event{} \models_\eta C$} \} \\ 
\end{align*}
\vspace{-12mm}

\begin{align*}                   
\delta_i^{loop}   & = \{ ((Q_i, \env, \envhat, \lev, \ltm), (\event{}, \timeof{}), (Q_i, \env, \envhat, \lev, \ltm)) \\
                  &  \text{ for every \env, \envhat, \lev, \ltm, \event{}, and \timeof{}} \\
                  & \text { unless $\event{} \models_\eta C$ or (\event{}=TICK \text{ and } $\timeof{} = \ltm+D$)} \\
\delta_i^{ok}     & = \{ ((Q_i, \env, \envhat, \lev, \ltm), (\event{}, \timeof{}), (Q_{i+1}, \env+[e, \event{}], \emptyset, \event{}, \timeof{})) \\
                  & \text{ where } \event{}=TICK \text{ and } \timeof{} = \ltm+D \}  \\
\end{align*}
\vspace{-11mm}

\begin{align*}                   
\delta_i^{loop}   & = \{ ((Q_i, \env, \envhat, \lev, \ltm), (\event{}, \timeof{}), (Q_i, \env, \envhat, \lev, \ltm)) \\
                  &  \text{ for every \env, \envhat, \lev, \ltm, \event{}, and \timeof{}, } \text{ unless $\event{} \models_\eta A$} \} \\
\delta_i^{block}  & = \{ ((Q_i, \env, \envhat, \lev, \ltm), (\event{}, \timeof{}), (Q_i, \env, \envhat', \lev, \ltm)) \\
                  &  \text{where $\event{} \models_\eta A-[e2]$, and } \envhat' = \envhat \cup \{ \emptyset+[e2, \event{}] \}  \} \\
\delta_i^{ok}     & = \{ ((Q_i, \env, \envhat, \lev, \ltm), (\event{}, \timeof{}), (Q_{i+1}, \env+[e2, \event{}], \emptyset, \event{}, \timeof{})) \\
                  & \text{ where $\event{} \models_\eta B$ and} \\
                  & \forall \env_{block} \in \envhat \,|\, \exists \mathtt{f} \in \mathtt{FIELDS} \,|\, getf(e2, \phi(f)) \neq \env_{block}(\mathtt{e2.f}) \\
\end{align*}
\vspace{-12mm}

\end{minipage}
\caption{\small Producing the non-deterministic transition relation for a query $\lbrack O_1, ..., O_n \rbrack$.
For each $i$, $\delta_i$ equals the union of all $\delta_i^{x}$.  
The notation $A-[e]$ means the subset of match predicate $A$ with all literals involving
event $e$ removed, and $\phi$ is the mapping on fields induced by equalities in the
blocking condition of an \qry{until}. E.g., in Property 4, \qry{nwSrc} maps to \qry{nwDst}
and vice versa.}
\label{fig:automata-full}
\hrule
\end{figure*}

\Cref{fig:automata-full} gives the complete transition-function definition (\Cref{sec:automata}).

\newpage
\section{Proof Sketch for Theorem 1}
\label{app:thm1}

% Renumber this theorem 1 by just creating a new counter
\spnewtheorem{rpttheorem1}{Theorem}{\bfseries}{\itshape}
\begin{rpttheorem1}
	Let $T$ be a trace and $Q = \lbrack O_1, ..., O_n \rbrack$ a query. Then $T \models Q \iff \queryaut$ accepts $T$.
\end{rpttheorem1}

\spnewtheorem*{proofsketch}{Proof Sketch}{\bfseries}{\itshape}
\begin{proofsketch}
	
	Note that $\queryaut$ accepts $T$ of length $k$ if and only if there is some run
	$(O_0, \emptyset, \emptyset, \bot, 0) \xrightarrow{\event{1}, \timeof{1}} ... \xrightarrow{\event{k}, \timeof{k}} (O_n, F_{\env}, F_{\envhat}, F_{\lev}, F_{\ltm})$
	of $\queryaut$ for some $\env$, $\envhat$, $\lev$ and $\ltm$. 
	
	For the $\leftarrow$ direction, if there is an execution of
	$\queryaut$ accepting $T$ then there is a partition of $T$ into
	non-empty traces $t_1, ..., t_n \; (t_1 \circ ... \circ t_n = T)$ such
	that the final event in each $t_i$ coincides with a transition from
	some $(O_i, \env, \envhat, \lev, \ltm)$ state to an $(O_{i+1}, \env', \envhat', \lev', \ltm')$ 
	state in the accepting execution. It suffices to show that, for each 
	$t_i$, $t_i \circ ... \circ t_n \models_{\eta, \ltm, \lev} \lbrack O_{i+1}, ..., O_n \rbrack$. 
	This holds by construction of $\queryaut$.
	
	The $\rightarrow$ direction proceeds similarly; given $T \models Q$ we build a set
	of sub-traces induced by a valid selection of indices ($i$ in \Cref{fig:semantics}) and
	produce an accepting execution.
\end{proofsketch}

\section{Compiler Algorithm}
\label{app:alg}

The query-compilation algorithm appears in full below. As seen in \Cref{sec:compile}, 
the compiler recursively descends through each observation,
building a corresponding set of switch rules.
Where match predicates involve negative literals, the compiler inserts
\emph{shadow} rules (\qry{buildShadow}) before the \emph{trigger} rule (\qry{buildTrigger})
that advances the state. Since only one rule per table
can apply, these block the trigger from applying.
The \qry{buildUntilBlock} function produces rule that learns new 
shadow rules that prevent the observation from
being triggered as blocking events
are seen. 

In the algorithm, \qry{Match} and \qry{Rule} are constructors that 
generate new match criteria and \ofpp rules. Each observation contains a \qry{type} field (ingress or egress), a matching predicate (i.e., list of literals) \qry{pred}, and in the case of \qry{until} observations, a blocking predicate \qry{blockPred}. 
\qry{NEXT(o.type)} denotes learning into the next unused table for \qry{o}'s type.
This value starts at 0 and is incremented by the \qry{inc} action.
\qry{SAME} indicates the rule is installed into the same table as its parent.

\begin{lstlisting}[label=lst:alg-buildShadow,language=query,numbers=none]
*{\bf\ttfamily compile(o: Observation, rest: List[Observation], depths: Map[Var, Int]):}*                      
  ss := buildShadow(o.cond, depths)            
  t  := buildTrigger(o, rest, depths)          
  if o.isUntil then                      
    bs := buildUntilBlock(o, depths)     
  else                                   
    bs := {}                             
  return (ss + t + bs)                   

*{\bf\ttfamily buildShadow(p: List[Literal], depths: Map[Var, Int]):}*
  ruleset := {}
  for each (t1 != t2) in pred:
    ruleset += Rule(o.time, NEXT(o.type), Match(t1, t2, depths), drop, None)
  return ruleset

*{\bf\ttfamily buildTrigger(o: Observation, rest: List[Observation], depths: Map[Var, Int]):}*
  matches := {}
  for each (t1 = t2) in o.pred:
    matches += Match(t1, t2, depths)
  
  if(rest.empty)
    actions  := {controller}
  else
    newrules := compile(rest.first, rest.rest, depths+{o.var->d.max})
    actions  := {inc(o.type), learn(newrules)}
    
  if(o.positive):
    return Rule(o.time, NEXT(o.type), matches, actions, None)
  else:
    return Rule(o.time, NEXT(o.type), matches, {delete_all}, actions) 

*{\bf\ttfamily buildUntilBlock(o: Observation, depths: Map[Var, Int]):}*             
  fieldMap := buildFieldMap(o.blockPred) // *{\rmfamily buildFieldMap (not shown) builds the relation}*
  ss := buildShadow(o.block, depths)     // *{\rmfamily $\phi$ in \Cref{fig:automata}. E.g., in Property 4,}* 
                                         // *{\rmfamily nwSrc maps to nwDst and vice versa.}* 
  restricted := o.blockPred - {literals using trigger var}
  matches := {}
  for each literal (t1 = t2) in restricted:
    matches += Match(t1, t2, depths)
    
  blockmatch := {}
  for each entry (fld1 -> fld2) in fieldMap:
    blockmatch += Match(fld2, fld1, depths)
  actions := {learn(Rule(o.time, SAME, blockMatch, drop, None))}
  
  t := Rule(o.time, SAME, matches, actions)	  
  return (ss + t)
\end{lstlisting}

\end{document}